\documentclass[twocolumn,showpacs,preprintnumbers,amsmath,amssymb]{revtex4-1}
\usepackage{dcolumn}
\usepackage{bm}
\usepackage[english]{babel}
\usepackage{graphicx}
\usepackage{graphics}
\usepackage{slashed}
\graphicspath{{FIGURE/}}

\begin{document}

\title{A chiral quark-soliton model with broken scale invariance for nuclear matter}

\author{Alessandro Drago and Mantovani Sarti Valentina}

\affiliation{Dipartimento di Fisica, Universita' di Ferrara and INFN, Sez. Ferrara, $44100$ Ferrara, Italy}

\begin{abstract}

We present a model for describing nuclear matter at finite density
based on quarks interacting with chiral fields, $\sigma$ and $\pi$
and with vector mesons introduced as massive gauge fields.
The chiral Lagrangian includes a logarithmic potential, associated
with the breaking of scale invariance. We provide results for the
soliton in vacuum and at finite density, using the Wigner-Seitz
approximation. We show that the model can reach higher densities
respect to the linear-$\sigma$ model and that the introduction of vector
mesons allows to obtain saturation. This result was never obtained
before in similar approaches.
 
\end{abstract}


\maketitle

\section{Introduction}

The problem of studying nuclear matter with chiral Lagrangians is not
trivial; for instance models based on the linear $\sigma$-model fail
to describe nuclear matter already at $\rho \sim \rho_0$ because the
normal solution in which chiral symmetry is broken becomes unstable
respect to the Lee-Wick phase. In Ref.~\cite{R.J.Furnstahl1993} 
Furnstahl and Serot conclude that the failure of the $\sigma$-model is due to the
restrictions on the scalar field dynamics imposed by the Mexican hat
potential. A possible solution to this problem is still to use a linear realization, but
with a new potential, which includes terms not present in the Mexican
hat potential.  A guideline in building such a potential is scale
invariance~\cite{E.K.Heide1994, G.W.Carter1998, G.W.Carter1997, G.W.Carter1996}.

In QCD the invariance under dilatation is spontaneously broken due to the presence of
the parameter $\Lambda _{QCD}$ coming from the renormalization
process.  Formally, the non conservation of the dilatation current is
strictly connected to a not vanishing gluon 
condensate:
\begin{equation}\label{current}
\langle \partial_{\mu}j^{\mu}_{QCD} \rangle =\dfrac{\beta(g)}{2g}\langle F^{a}_{\mu \nu}(x)F^{a\mu \nu}(x)\rangle .
\end{equation}

In the approach of Schechter~\cite{J.Schechter1980}, and of Migdal and Shifman~\cite{Migdal:1982jp}
 a scalar field representing the gluon
condensate is introduced and its dynamics is regulated by a potential
chosen so that it reproduces (at Mean-Field level) the divergence of
the scale current that in QCD is given by Eq.~(\ref{current}). The
potential of the dilaton field is therefore determined by the
equation:
\begin{equation}\label{dil}
\theta_\mu ^\mu = 4 V(\phi)-\phi \dfrac{\partial V}{\partial \phi}=4
\epsilon_{vac} \left (\dfrac{\phi}{\phi_0}\right )^4
\end{equation}
 where the parameter $ \epsilon_{vac}$ represents the vacuum energy.
To take into account massless quarks a generalization was proposed in
Ref.~\cite{E.K.Heide1992}, so that also chiral fields contribute to
the trace anomaly. In this way the single scalar field of
Eq.~(\ref{dil}) is replaced by a set of scalar and pseudoscalar fields $\lbrace \sigma,
\boldsymbol{\pi}, \phi \rbrace$.

It has already been shown that an hadronic model based on this
dynamics provides a good description of nuclear physics at densities
about $\rho_0$ and it describes the gradual restoration of chiral
symmetry at higher densities~\cite{Bonanno:2008tt}. In the same work
the authors have shown a phase diagram, where the interplay between chiral
and scale invariance restoration lead to a scenario similar to that
proposed by McLerran and Pisarski in~\cite{L.McLerran2007}. 
This is not too surprising since the large $N_c$ limit explored in
\cite{L.McLerran2007} should be well represented in chiral models as the
one discussed in \cite{Bonanno:2008tt}.
It is therefore tempting to explore the scenario presented in
~\cite{Bonanno:2008tt} at a more microscopic level.

\textit{The new idea we develop in this work is to interpret the
fermions as quarks, to build the hadrons as solitonic solutions of the
fields equations as in~\cite{Birse:1986qc, Kahana:1984be} and, finally,
to explore the properties of the soliton at finite density using the
Wigner-Seitz approximation.}

Similar approaches to a finite density system have been investigated
in the past~\cite{Achtzehnter:1985ur, N.K.Glendenning1986, D.Hahn1987,
Birse:1987pb, Birse:1991cx, U.Weber1998, P.Amore2000}.  A problem of
those works is that the solitonic solutions are unstable and disappear
already at moderate densities when e.g. the linear $\sigma$-model is
adopted~\cite{U.Weber1998}.  We are therefore facing an instability
similar to the one discussed and solved when studying nuclear matter
with hadronic chiral Lagrangians.
The first aim of our work is \textit{to check whether the inclusion of the
logarithmic potential allows the soliton crystal to reach higher
densities.} The second and more important aim is \textit{to check
whether the inclusion of vector mesons in the dynamics of the quarks
can provide saturation for chiral matter.}

We should remark that the version of this model without
vector mesons has already been studied
at zero density in~\cite{AbuShady:2010zza}, but using a different
technique to describe the single nucleon. Instead no calculation
exists with the vector mesons. Moreover, both versions of the model
are unexplored at finite density.

The structure of the paper is as follows. In Sec.~\ref{model} we
describe the model we are using, in Sec.~\ref{hedgehog} we review the
Mean-Field approximation and the hedgehog ansatz while in Sec.~\ref{proiezione}
we present the technique adopted for projection.
Later in Sec.~\ref{hedgedensity} we describe the Wigner-Seitz approximation used to mimic 
a system at finite density.
 Next in
Sec.~\ref{resultsvac} and in Sec.~\ref{hedgefindensity} we present our results, firstly for the soliton in vacuum  
both at
Mean-Field level and by adopting a projection technique and then
 we show the results for the
Wigner-Seitz lattice of solitons. Finally, in Sec.~\ref{conclusions}
we present our conclusions and future outlooks.

\section{The model}\label{model}

In a chiral quark-soliton model quarks are coupled to mesons in a chirally invariant way. 
Following Refs.~\cite{E.K.Heide1994,G.W.Carter1998,G.W.Carter1997,G.W.Carter1996,Bonanno:2008tt} 
we consider the Lagrangian:
\begin{align}\label{lagr}
\mathcal{L}_0  =& \bar{\psi}  \big(i \gamma ^\mu\partial _{\mu}-
g_\pi(\sigma + i \boldsymbol{\pi}\cdot \boldsymbol{\tau}\gamma _{5}))\psi \nonumber\\
  &+\dfrac{1}{2}(\partial_{\mu}\phi \partial^{\mu}\phi +
\partial_{\mu}\sigma \partial^{\mu}\sigma + \partial_{\mu}\boldsymbol{\pi}\cdot \partial^{\mu}\boldsymbol{\pi})\nonumber\\
  &-V(\phi,\sigma,\mathbf{\pi}).
\end{align}
Here $\psi$ is the quark field, $\sigma$ and $\pi$ are the chiral fields and $\phi$ is
the dilaton field which, in the present calculation, is kept frozen at its vacuum value $\phi _0$.

An extension of this model,
already discussed in Refs.~\cite{E.K.Heide1994,G.W.Carter1998,G.W.Carter1997,G.W.Carter1996,Bonanno:2008tt},
 is to add the dynamics of vector mesons 
and to incorporate the idea of universal coupling~\cite{Sakurai1966}. 
This can be achieved by considering the vector mesons as massive gauge fields,
following also the scheme of Ref. \cite{W.Broniowski1986}.
The new Lagrangian is given by:
\begin{align} \label{lagrVM}
\mathcal{L}_{VM} & =\bar{\psi}  \big(i \gamma ^\mu\partial _{\mu}-g_{\pi}(\sigma + i \boldsymbol{\pi}\cdot 
\boldsymbol{\tau}\gamma _{5}) +g_{\rho}\gamma ^\mu \dfrac{\boldsymbol{\tau}}{2}\cdot 
(\boldsymbol{\rho} _{\mu}+\gamma _{5} \boldsymbol{A} _{\mu})\nonumber \\
& -\dfrac{g_{\omega}}{3} \gamma ^\mu \omega _{\mu}\big)\psi 
  +\dfrac{\beta}{2}(D_{\mu}\sigma D^{\mu}\sigma + D_{\mu}
  \boldsymbol{\pi}\cdot D^{\mu}\boldsymbol{\pi})\nonumber \\
& -\dfrac{1}{4}(\boldsymbol{\rho}_{\mu \nu}\cdot \boldsymbol{\rho}^{\mu \nu}+
\boldsymbol{A}_{\mu \nu}\cdot \boldsymbol{A}^{\mu \nu} + \omega_{\mu \nu} \omega ^{\mu \nu})\nonumber \\
&   +\dfrac{1}{2}m_{\rho}^2(\boldsymbol{\rho}_{\mu} \cdot \boldsymbol{\rho}^{\mu}
+\boldsymbol{A}_{\mu} \cdot \boldsymbol{A}^{\mu})+\dfrac{1}{2}m_\omega ^2 \omega _\mu \omega ^\mu \nonumber\\
& -V(\phi_0,\sigma,\mathbf{\pi})
\end{align}
where $\omega_\mu$ is a vector-isoscalar coupled to baryon current,
 $\boldsymbol{\rho}_\mu$ and $\boldsymbol{A}_\mu$ are respectively a vector-isovector
  and an axial-vector- isovector fields coupled to isospin and axial-vector current.
The covariant derivatives for the chiral fields and the field tensors for vector mesons read:
\begin{align}
&D_\mu \sigma =\partial_\mu \sigma + g_\rho \boldsymbol{A}_\mu \cdot \boldsymbol{\pi}\nonumber ,\\
&D_\mu \boldsymbol{\pi} =\partial_\mu \boldsymbol{\pi} + 
g_\rho (\boldsymbol{\rho}_\mu \wedge \boldsymbol{\pi} - \boldsymbol{A}_\mu \sigma)\nonumber ,\\
 &\omega_{\mu\nu} = \partial_\mu \omega_\nu-\partial_\nu \omega_\mu \nonumber,\\ &\boldsymbol{\rho}_{\mu\nu}=\partial_\mu \boldsymbol{\rho}_\nu-\partial_\nu \boldsymbol{\rho}_\mu+
 g_\rho(\boldsymbol{\rho}_\mu \wedge \boldsymbol{\rho}_\nu + \boldsymbol{A}_\mu \wedge \boldsymbol{A}_\nu)\nonumber ,\\
& \boldsymbol{A}_{\mu\nu}=\partial_\mu \boldsymbol{A}_\nu-
\partial_\nu \boldsymbol{A}_\mu + g_\rho(\boldsymbol{\rho}_\mu \wedge \boldsymbol{A}_\nu + \boldsymbol{A}_\mu \wedge \boldsymbol{\rho}_\nu).
\end{align}
The pion field mixes with the longitudinal component of the $a_1$~\cite{Gasiorowicz:1969kn, Ko:1994en}
 and so to ensure that the pion gets its physical mass once the two fields are decoupled, we need to introduce the constant:
\begin{equation}\label{beta}
\beta= \dfrac{m_\rho ^2}{m_\rho ^2 -g_\rho ^2 f_\pi ^2}.
\end{equation}

The potential is given by:
\begin{align} \label{pot}
V(\phi,&\sigma,\mathbf{\pi}) =  B\phi_0 ^4\left( \ln \dfrac{\phi}{\phi_0} -\dfrac{1}{4}\right ) -
\dfrac{1}{2}B \delta \phi^4 \ln \dfrac{\sigma^2 + 
\boldsymbol{\pi}^2}{\sigma_{0} ^2}\nonumber\\
  +& \dfrac{1}{2}B\delta \phi^2\dfrac{\phi_0 ^2 }{\sigma^2_0} 
\left (\sigma^2 + \boldsymbol{\pi}^2-\phi^2 \dfrac{\sigma^2_0}{2 \phi_0 ^2}\right ) \\
  -&  \dfrac{1}{4} \epsilon_{1} \left (\dfrac{\phi}{\phi_0}\right)^2 \left[\dfrac{4\sigma}{\sigma _{0}}
 -2 \left (\dfrac{\sigma^2 + \boldsymbol{\pi}^2}{\sigma_0^2}\right)-
\left(\dfrac{\phi}{\phi_0}\right)^2\right]-V_0\nonumber
\end{align}
where the logarithmic term generates from~(\ref{dil}). The first two
terms of the potential are responsible for the breaking of scale
invariance, while the second line is needed to ensure that in the
vacuum $\phi=\phi_0$, $\sigma=\sigma_0$ and $\boldsymbol{\pi}=0$, i.e
it provides spontaneous symmetry breaking. The last line
explicitly breaks the chiral invariance of the lagrangian, giving mass to the pion.

In this work we assign the following values to the masses of
 bare fields: $m_\pi =139$ MeV, $m_\rho=m_A=776$ MeV and $m_\omega=782$ MeV. 
 For the sigma field, since there are no experimental constraints, 
 we use $m_\sigma =550$ MeV and $m_\sigma=1200$ MeV which are typical values used in nuclear physics
if the sigma has to provide the intermediate range attraction.
We relate the quark-omega and the quark-rho couplings to the corresponding
couplings with the nucleons. We keep fixed $g_\rho=4$ and we vary $g_\omega$ 
between $10$ and $13$. The pion-quark coupling constant $g_\pi$
will vary from $3.9$ to $5$.
By varying the coupling constants in the ranges indicated above we will
be able to tune the attractive and the repulsive interactions, in order to
obtain saturation when studying the total energy of the soliton at finite 
density.
The constants $B$ and $\phi_0$ can be fixed by choosing a value for the
mass of the glueball and for the vacuum energy $\epsilon_{vac}$, while
$\delta=4/33$ is provided by the QCD beta function and it corresponds
to the relative weight of the fermionic and of the gluonic degrees of
freedom.  Finally the constant $V_0$ ensures that the potential energy
is vanishing in the vacuum.

As anticipated, we choose to keep the dilaton field frozen.
This decision (which obviously
simplifies the dynamics of the system) can be justified by the results
obtained in~\cite{Bonanno:2008tt, G.W.Carter1998}, where it has been
shown that at low temperatures the dilaton remains close to its vacuum
value even at large densities.  Therefore the degrees of freedom of
our model are limited to quarks, chiral fields and vector mesons, similarly to the
linear $\sigma$-model whose solitonic solutions have been obtained
in~\cite{Birse:1986qc,W.Broniowski1986}. The potential can be written
in a simpler form and it reads:
\begin{align} \label{potfro}
& V(\sigma,\mathbf{\pi}) =  \lambda_1 ^2 (\sigma^2 +\boldsymbol{\pi}^2)-\lambda_2 ^2\,\ln(\sigma^2 + 
\boldsymbol{\pi}^2)-\sigma_0 m_\pi ^2 \sigma
\end{align}
where:
\begin{align}
& \lambda_1 ^2=\dfrac{1}{2}\dfrac{B \delta \phi_0 ^4+\epsilon_1}{\sigma_0 ^2}=\dfrac{1}{4}(m_\sigma ^2+m_\pi ^2)\\
& \lambda_2 ^2=\dfrac{1}{2}B \delta \phi_0 ^4= \dfrac{\sigma_0 ^2}{4}(m_\sigma ^2-m_\pi ^2)\\
& \epsilon_1 =m_\pi ^2 \sigma_0 ^2.
\end{align}
Here the vacuum value $\sigma_0$ is fixed to be equal to $f_\pi =93$
MeV. 

\begin{figure}[t]
\centering
\includegraphics*[width=\columnwidth]{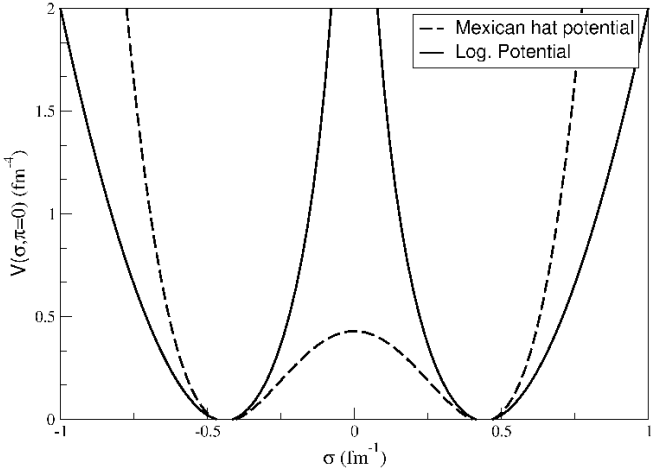} 
\caption{Comparison between the logarithmic (solid line) and the Mexican hat potential (dashed line).}\label{comppot}
\end{figure}

It is interesting to compare the logarithmic with the Mexican hat
potential. In Fig.~\ref{comppot} it can be seen that in the case of the
Mexican hat potential it is relatively easy to restore chiral symmetry
by climbing the maximum located at the center.  This is not
possible in the case of the logarithmic potential as long as the
dilaton field remains frozen. Since only at large temperatures the
dilaton field changes significantly~\cite{Bonanno:2008tt}
we can expect that at large
densities and moderate temperatures this model provides more stable
solitonic solutions.
This is a crucial question which will be investigated in our paper.

An important point in our approach is that we aim at describing
the dynamics of nuclear matter by incorporating all the interactions
already at a quark level. This is at variance with e.g. the
approach of Ref.\cite{Smith:2003hu, PhysRevC.70.065205} where the vector field $\omega$ 
was introduced only at the nucleon level, but was not
present in the dynamics of the quarks.

\section{Mean-Field and hedgehog ansatz}\label{hedgehog}

The Euler-Lagrangian equations that follow from the Lagrangian 
given in eq.~(\ref{lagrVM}) read:

\begin{align}
& [i\slashed\partial-g_{\pi}(\sigma + i \boldsymbol{\pi}\cdot \boldsymbol{\tau}\gamma _{5}) 
+g_{\rho} \dfrac{\boldsymbol{\tau}}{2}\cdot (\boldsymbol{\slashed\rho} 
+\gamma _{5} \boldsymbol{\slashed A} )-\dfrac{g_{\omega}}{3} \slashed\omega ]\psi = 0 \nonumber,\\
& \beta \partial _\mu D^\mu \sigma = -\beta g_\rho \boldsymbol{A} _{\mu}\cdot \boldsymbol{\pi}
-g\bar{\psi}\psi -\dfrac{\partial V}{\partial \sigma}\nonumber,\\
& \beta \partial _\mu D^\mu \boldsymbol{\pi} = \beta g_\rho (-\boldsymbol{\rho} _{\mu} \times D^\mu \boldsymbol{\pi}
+\boldsymbol{A} _{\mu}D^\mu \sigma)-i g \bar{\psi}\boldsymbol{\tau}\gamma_5 
\psi-\dfrac{\partial V}{\partial \boldsymbol{\pi}}\nonumber,\\
& -\partial ^\mu \boldsymbol{\rho}_{\mu \nu}=g_\rho \boldsymbol{\upsilon}_\nu 
+m_\rho ^2\boldsymbol{\rho}_{\nu}\nonumber ,\\
& -\partial ^\mu \boldsymbol{A}_{\mu \nu}=g_\rho \boldsymbol{a}_\nu 
+m_\rho ^2\boldsymbol{A}_{\nu}\nonumber ,\\
& -\partial ^\mu\omega_{\mu \nu}=-\dfrac{1}{3}g_\omega \bar{\psi}\gamma_\nu\psi +m_\omega ^2\omega_{\nu}.
\end{align}
Here  $\boldsymbol{\upsilon}_\nu$ and $\boldsymbol{a}_\nu$ are the vector and the axial-vector currents:
\begin{align}
&\boldsymbol{\upsilon}_\nu = \boldsymbol{\rho}_{\mu}\times \boldsymbol{\rho}_{\mu \nu}
+\boldsymbol{A}_{\mu}\times \boldsymbol{A}_{\mu \nu}+
\beta \boldsymbol{\pi}\times D^\nu \boldsymbol{\pi}+\bar{\psi}\gamma_\nu \dfrac{\boldsymbol{\tau}}{2}\psi,
\end{align}
\begin{align}
\boldsymbol{a}_\nu = \boldsymbol{\rho}_{\mu}\times & \boldsymbol{A}_{\mu \nu}
+\boldsymbol{A}_{\mu}\times \boldsymbol{\rho}_{\mu \nu}+\beta \boldsymbol{\pi}\times D^\nu \sigma-\beta\sigma D_\nu \boldsymbol{\pi}\nonumber\\
& +\bar{\psi}\gamma_5\gamma_\nu \dfrac{\boldsymbol{\tau}}{2}\psi.
\end{align}

The fields equations for the model without vector mesons, provided by the Lagrangian in eq.~(\ref{lagr}),
 can be obtained by the previous ones by fixing $g_\rho=g_\omega=0$.\\ 

The previous equations are relations between quantum fields.
The starting point of our calculation is the Mean-Field approximation,
where mesons are described by time-independent, classical fields and
where powers and products of these fields are replaced by powers and
products of their expectation values.  The quark spinor in the
spin-isospin space is:
\begin{equation}
\psi=\dfrac{1}{\sqrt{4 \pi}}\left (\begin{array}{c}
u(r)\\
iv(r)\boldsymbol{\sigma} \cdot \boldsymbol{\hat r}\end{array}\right) \chi_h
\end{equation}
where the spinor $\chi_h$,
defined as:
\begin{equation}
\chi_h= \dfrac{1}{\sqrt 2}(|u_\downarrow\rangle-|d_\uparrow\rangle)
\end{equation}
 satisfies the condition for the Grand Spin $\boldsymbol{G}=\boldsymbol{I}+\boldsymbol{J}$:
\begin{equation}\label{grandspin}
\boldsymbol{G}\chi_h=0.
\end{equation}
The hedgehog baryon $\vert B\rangle =\vert N_c q\rangle \vert \sigma
\rangle \vert \pi \rangle\vert \omega\rangle \vert \rho\rangle \vert A\rangle$ is given by the product of quarks and
coherent states of mesons fields and it corresponds to a linear
combination of states with $I=1/2$ and $I=3/2$:
\begin{equation}\label{hedgestate}
\vert B\rangle = \sum_{JMM_I}(-)^{J+M}C_J\delta_{M,-M_I}\vert J=I,M=-I_3,M_I\rangle.
\end{equation}
An explicit expression for the coefficients $C_J$ will be given in Sec.~\ref{proiezione}.

It can be shown that the hedgehog state is one element 
of an infinite class of degenerate solutions of field equations at Mean-Field level~\cite{Fiolhais:1985bt}.
The profiles of the chiral and the vector mesons fields 
in the hedgehog state are given by the relations:
\begin{align}
& \dfrac{\langle B\vert \sigma \vert B\rangle}{\langle B\vert B\rangle} =
\sigma_h (r)\nonumber ,\quad \dfrac{\langle B\vert\boldsymbol{\pi}_a \vert B\rangle}
{\langle B\vert B\rangle} =\widehat{r}_a h (r)\nonumber\\
& \dfrac{\langle B\vert \omega_\mu \vert B\rangle}{\langle B\vert B\rangle}
\omega_0 (r)=\omega (r)\nonumber\nonumber\\
& \dfrac{\langle B\vert\boldsymbol{\rho}_a \vert B\rangle}
{\langle B\vert B\rangle} =\rho_i ^a(r)=\rho (r)\epsilon^{ika} \widehat{r}^k\nonumber\\
& \dfrac{\langle B\vert\boldsymbol{A}_a \vert B\rangle}
{\langle B\vert B\rangle} =A_i ^a(r)=A_S (r)\delta ^{ai}
+A_T (r)(\widehat{r}^a \widehat{r}^i-\dfrac{1}{3}\delta^{ai})\nonumber
\end{align}
At Mean-Field level the mesons fields are classical and the
differential equations governing their dynamics have to be
supplemented by the appropriate boundary conditions.  For the single
nucleon case we impose the following boundary conditions to the
fields:
\begin{eqnarray}\label{bcvac}
& u'(0)=v(0)=0\nonumber ,\\
&\sigma_h '(0)=h (0)=0,\\
&\rho (0)=\omega '(0)=A_S '(0)=A_T (0)=0, \nonumber
\end{eqnarray}
while at infinity the boundary conditions read:
\begin{align}
& \sigma_h (\infty)=\sigma_0 ,\,  h (\infty)=0,\nonumber\\
& \dfrac{v(\infty)}{u(\infty)}=\sqrt{\dfrac{-g \sigma_0+\varepsilon}{-g \sigma_0-\varepsilon}} , \,\\ 
& \omega'(\infty)=\rho '(\infty)=A_S '(\infty)=A_T '(\infty)= 0,\nonumber
\end{align}
where $\varepsilon$ is the quark eigenvalue.\\
The total energy of the soliton at Mean-Field level is given by:
\begin{align}\label{enmfa}
 E_{MFA}= 4\pi \int &r^2 dr (E_{int}+E_{kin,Q}+E_{\sigma}+E_\pi \nonumber\\
 & +E_\omega +E_\rho +E_A+ E_{pot})
\end{align}
where explicit expressions for each term in the energy density can be found in Appendix I. 
A test of the numerical accuracy of the solution originates from
another way of expressing the energy, obtained by
Rafelski~\cite{Rafelski1977} by integrating out the fermionic fields:
 \begin{align}
E_{Raf.}&=\int  d^3r\bigg[4\left(V-\sigma\dfrac{\partial V}{\partial \sigma}-\pi \dfrac{\partial V}{\partial \pi}\right)-m_\omega ^2\omega ^2 \nonumber\\
& +m_\rho ^2\left(2\rho ^2+3A_S ^2+\dfrac{2}{3}A_T ^2\right)\bigg]
\end{align}
Our solutions satisfy this consistency test up to a precision of the order of $10^{-3}$.


\section{Projection}\label{proiezione}

The hedgehog baryon defined in section~\ref{hedgehog} is not an
eigenstate of isospin and angular momentum. Moreover this
semi-classical solution also breaks the translational symmetry of the
Lagrangian, since the localized soliton is not an eigenstate of the 
linear momentum, either. In this work we restore the
invariance under rotations by using the projection technique developed
in~\cite{Ruiz1995,Birse:1986qc}.  Instead, we will not restore the
translational invariance of our soliton. The spin-isospin eigenstates
are defined as follows:
\begin{equation}\label{projstate}
\vert JMM_I\rangle = N_{JM_I} \int d^3\Omega D_{M,-M_I}^{J}(\Omega)^* \widehat{R}(\Omega)\vert B\rangle
\end{equation}
where the weight functions $D$ are the Wigner functions,
$\widehat{R}(\Omega)$ is a spatial rotation through Euler angles
$\Omega \equiv (\alpha,\beta,\gamma)$ and $ N_{JM_I}$ is a
normalization factor.\\
Since the hedgehog states are eigenstates of the Grand Spin $\boldsymbol{G}$, 
it is equivalent to rotate either in spin or in isospin space.
Moreover, when studying diagonal matrix elements of nucleon states, it
is customary to work with states where the third component of the
angular momentum $M$ is equal to $-M_I$ since in this case the expression of
the Wigner function is particularly simple.  In this way the
projection operator becomes:
\begin{equation}\label{projoperator}
P_{JM}=\dfrac{2J+1}{8\pi^2}\int d^3 \Omega D_{M,M}^{J}(\Omega)^* \widehat{R}(\Omega).
\end{equation}
The normalization factor has been determined by using~(\ref{projstate})-(\ref{projoperator}):
\begin{align}
N^2 _{J, -M} & = \left(\dfrac{2J+1}{8\pi^2}\right)^2 (\langle B \vert P_{JM}\vert B\rangle)^{-1}.
\end{align}
Finally, the coefficients $C_J$ in eq.~(\ref{hedgestate}) are given by the expression:
\begin{align}
C_J ^2 & = \langle B \vert P_{JM}\vert B\rangle\nonumber \\
& = \dfrac{2J+1}{8\pi^2}\int d^3\Omega  D_{M,M}^{J}(\Omega) \langle B\vert\widehat{R}(\Omega)\vert B\rangle.
\end{align}
Once we obtain the projected state, we proceed to evaluate the corresponding energy.
Basically we need to calculate the expectation value of the Hamiltonian on the projected state given by~(\ref{projstate}).
The projected energy can be written as:
\begin{align}\label{projen}
& E_{J} & = &\langle JM-M\vert :H:\vert JM-M\rangle \nonumber\\
& & = &4\pi \int r^2 dr (E_{int}+E_{kin,Q}+E_{J,\sigma}+E_{J,\pi}\nonumber\\
& & &+E_{\omega}+E_{J,\rho}+E_{J,A}+E_{J,pot})
\end{align}
More details about the projection technique and the evaluation of the each term in the energy can be found in~\cite{Birse:1986qc, Ruiz1995}.\\

\subsection{Projected observables}

The formalism needed to compute most of the observables can be found
in~\cite{Birse:1986qc,Ruiz1995}.  The only quantity for which we need
to provide a new explicit expression is the potential energy, since in our
potential there is a logarithmic term, not present in the $\sigma$
model.\\
The matrix element for which we need to develop a full calculation is:
\begin{equation}\label{potproj}
E_{J,pot}= \langle JM-M\vert :\int d^3r V(\sigma_h,h):\vert JM-M\rangle
\end{equation}
 More details about the calculation of this term are given in Appendix II.\\
To compute the static
observables we have used the explicit formulae given in the Appendix
of Ref.~\cite{Ruiz1995}; in addition to radii and magnetic moments we also
show the results for the average number of pions in the projected
state, given by:
\begin{equation}
\langle N_\pi\rangle _J=\overline{N}_\pi C_0 (J,\overline{N}_\pi),
\end{equation}
where $C_0(J,\overline{N}_\pi)$~\cite{Birse:1986qc} is a projection coefficient depending
on the spin and on the average number of pions $\overline{N}_\pi$ in
the unprojected state. 
As already mentioned before, we do not perform a
projection on the linear momentum, but we adopt an easier
approach~\cite{Dethier:1982ax} which provides a rough estimate of the
center-of-mass corrections to the baryon total energy.  The masses for
the $N$ $(J=1/2)$ and for the $\Delta$ $(J=3/2)$ are then given by:
\begin{equation}\label{spurious}
M_J = (E_J-\boldsymbol{P}^2)^{1/2} \, .
\end{equation}

%

\section{Wigner-Seitz approximation to nuclear matter}\label{hedgedensity}

In order to describe a soliton system at finite density we use the
Wigner-Seitz approximation. This approach is very common
and it has already been widely applied to both
non-linear~\cite{D.Hahn1987,N.K.Glendenning1986,P.Amore2000} and
linear-$\sigma$ models~\cite{U.Weber1998}. Specifically, the
Wigner-Seitz approximation consists of replacing the cubic lattice by
a spherical symmetric one where each soliton sits on a spherical cell
of radius R with specific boundary conditions imposed on fields at the
surface of the sphere.  The configuration of the meson fields,
centered at each lattice point, generates a periodic potential in
which the quarks move.

The spinor of quark fields
must satisfy the Bloch theorem:
\begin{equation}
\psi _{\boldsymbol{k}} (r)=e^{i \boldsymbol{k}\cdot \boldsymbol{r}}\Phi_{ \boldsymbol{k}}(r),
\end{equation}
where $\boldsymbol{k}$ is the crystal momentum (which for the ground state
is equal to zero) and $\Phi_{ \boldsymbol{k}}(r)$ is a spinor that has
the same periodicity of the lattice.

\subsection{Boundary conditions}
In the literature various sets of possible boundary conditions have
been discussed~\cite{U.Weber1998,P.Amore2000}.  In our work we adopt
the choice of Ref.~\cite{U.Weber1998} which relates the boundary conditions
at $R$ to the parity operation, $\boldsymbol{r}\rightarrow -\boldsymbol{r}$.
Respect to this symmetry the lower component $v(r)$ of quark spinor,
the pion $h (r)$ and the rho $\rho (r)$ are odd, and therefore they have to vanish at $R$:
\begin{equation}
v(R)=h (R)=\rho (R)=0.
\end{equation}
Similarly, for the $\sigma$ field, the upper Dirac component, the $\omega$ and the $A$ fields the 
argument based on parity provides the conditions:
\begin{equation}
u'(R)=\sigma_h '(R)=\omega '(R)=A_S '(R)=A_T '(R)=0.
\end{equation}

The boundary conditions at $r=0$ remain the ones given in eq.~(\ref{bcvac}).
Basically the calculation consists in solving the set of coupled field
equations in a self-consistent way for a given value $R$; practically we
start from $R=4$ fm, for which the periodic solutions are
indistinguishable from the vacuum ones, and we slowly decrease the
cell radius down to the smallest radius for which self-consistent
solutions can be obtained. The parameter set is the same used for
vacuum calculations.

\begin{figure}[t]
\centering
\includegraphics*[width=\columnwidth]{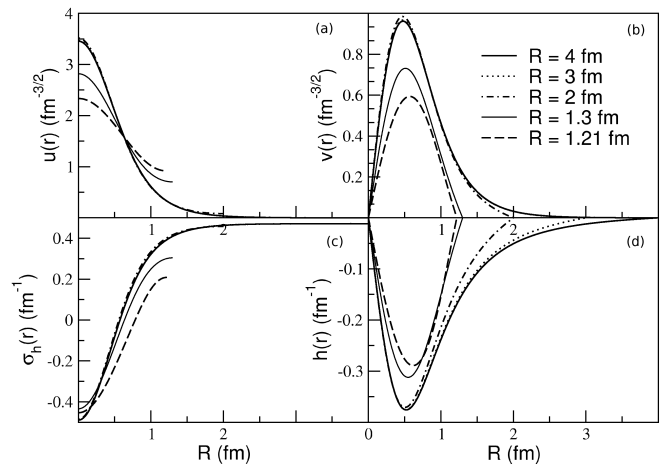}
\caption{Upper and lower components of the Dirac spinor ((a), (b)), $\sigma$ and 
pion fields ((c), (d)), in the model without vector mesons, as functions of the cell radius $R$.}\label{campi}
\end{figure}

In Fig.~\ref{campi} we plot the Dirac and the chiral fields in the model without vector mesons 
for different values of $R$; down to $R=2$ fm, the solutions do not change
significantly, but as the cell radius shrinks to lower values, we see
that all the fields are deeply modified by the finite density.

\begin{figure}[t]
\centering
\includegraphics*[width=\columnwidth]{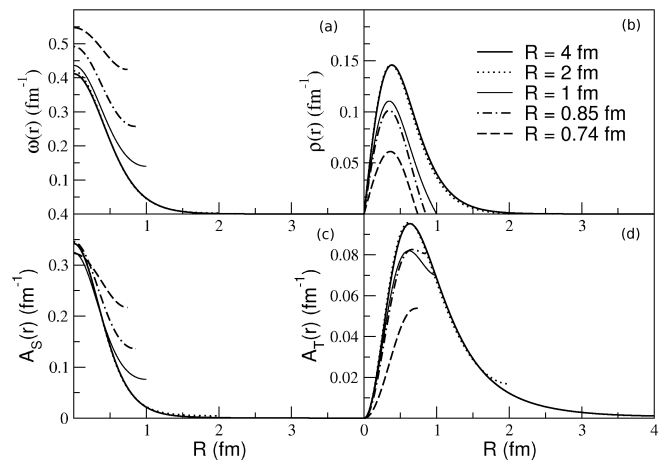}
\caption{Vector mesons fields, as functions of the cell radius $R$.
In (a) we plot the $\omega$ field, in (b) the $\rho$ field, in (c) the $A_S$ field and in (d) the $A_T$ field.}\label{campiVM}
\end{figure}

\begin{figure}[t]
\centering
\includegraphics*[width=\columnwidth]{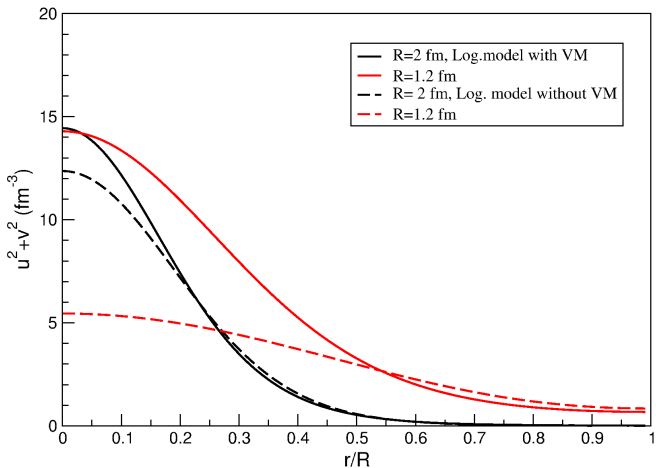}
\caption{Color online. Baryon density as a function of the ratio $r/R$ for the model without (dashed line) and with vector mesons (solid line). Two values of the cell radius $R$ are shown, namely $R=2$ fm and $R=1.2$ fm.
}\label{densityvm}
\end{figure}

For the model including vector mesons, we present in Fig.~\ref{campiVM} the trend of the vector 
mesons fields.
To better clarify the difference between the models without and with vector mesons, in Fig.~\ref{densityvm} we show the baryon density profiles in the two cases.
The relevant feature is that in the model without vector mesons the shape of the soliton becomes significantly more flat, at large densities, than in the case with vector mesons.
This effect is due to the repulsion between the two solitons provided by the $\omega$ field,
which prevents the baryon density to become large in the inter-nucleon region.
This feature will have an important consequence on the dependence of the radii on the density,
as discussed in Sec~\ref{properties}.
%

\subsection{Band width}
We are solving a problem in which quarks are moving in a
periodic potential and therefore Bloch theorem tells us that a band
should form. How to define the width of the band is highly not
trivial.  The most sophisticated technique is the one provided
in~\cite{U.Weber1998}.\\ 
Here we adopt two much simpler procedures. The first method is taken
from~\cite{D.Hahn1987}, where the authors estimates the band
width as:
\begin{eqnarray}\label{bandglend}
\Delta & =& \sqrt{\epsilon _0 ^2 + \left(\dfrac{\pi}{2R}\right)^2}-|\epsilon_0| , \\
\epsilon_{top} &=&\epsilon_0 +\Delta ,
\end{eqnarray}
where $\epsilon _0$ is the eigenvalue of the ground state.\\
An alternative approximation to the band width is obtained, following~\cite{Birse:1987pb}, 
by imposing that the lower Dirac component vanishes at the  boundary:
\begin{equation}\label{bandmike}
u(R)=0.
\end{equation}

A more accurate evaluation of the band effects can be obtained by
solving in a self-consistent way the field equations,which depend also
on $\boldsymbol{k}$, as discussed in~\cite{U.Weber1998}.  There it is
shown that $\epsilon_{top}$ obtained by imposing the condition~(\ref{bandmike}) is an upper limit to the top of the
band and that the true top lies about half way between this upper
limit and the bottom of the band. In conclusion, the band width turns
out to be considerably smaller than the one estimated using the first
method which leads to the upper limit $\epsilon_{top}$ in eq.~(\ref{bandglend}).

Concerning the filling of the band, when working with chiral solitons
at Mean-Field level the relevant quantum number is the grand-spin $G$
and the lower band corresponds to $G=0$. The only degeneracy remaining
is color and therefore the three quarks per soliton are completely
filling the band. The total energy of the cell is estimated by assuming a uniform
filling and by averaging the energy within the band.

\section{Results: properties of the nucleon in vacuum}\label{resultsvac}

We start by showing how the 
solitonic solutions for the
single nucleon can be built and we check
that it is possible to provide a reasonable description of the single
nucleon properties with the chosen parameter set.
In particular we present here results for the lagrangian without 
vector mesons $\mathcal{L}_0$ for the set $m_\sigma=550$ MeV and the
typical value $g_\pi=5$~\cite{E.K.Heide1992, Birse:1986qc}. Instead
the parameters for the model with vector mesons are given by $m_\sigma=1200$ MeV, $g_\pi=3.9$,
$g_\omega =12$ and $g_\rho =4$.
This set has been chosen in order to both
get saturation at finite density and a reasonable description of the
nucleon in vacuum. Anyway, since the calculation of nuclear matter properties
done by using the Wigner-Seitz approximation is affected by large uncertainties
we also present a second set of results with parameters better fitted to 
single nucleon properties.

In Tables~\ref{res} and~\ref{resVM} we present the static properties of the hedgehog
baryon at Mean-Field level and we compare them with results obtained
in the linear $\sigma$-model~\cite{Birse1985, W.Broniowski1986}. 

In Tables~\ref{contrib} and~\ref{contribVM}
we show the decomposition of the soliton total energy in its various
contributions and again we compare with the linear $\sigma$-model~\cite{Birse1985, W.Broniowski1986}.
We see that the contributions to the total energy coming from the chiral fields
and from the vector mesons are comparable: as expected vector mesons play an important role in the dynamics of the soliton.
It is also interesting to notice that the results obtained with the logarithmic model are very similar
to the ones obtained with the Mexican hat potential, at zero density, specially when vector mesons are included.

\begingroup
\begin{table}[t]
\caption{Various nucleon properties at Mean-Field level in the present work without vector mesons and in the $\sigma$-model~\cite{Birse1985}.}\label{res}
\centering
\begin{tabular}{|c | c  c  c|}
\hline 
 & & & \\
 Quantity & Log. Model & $\sigma$-Model & Exp.  \\[7pt]
 \hline
 & & & \\
 $M\,(MeV)$ & $1176$ & $1136$ & $1085$\\[5pt]
 $\langle r_e ^2\rangle_{I=0}$& $(0.76\, fm)^2$ & $(0.78\, fm)^2$  & $(0.72\, fm)^2$ \\ [5pt]
 $\mu_{I=1}$ $(\mu_N)$& $3.83$ & $3.63$ & $4.70$ \\[5pt] 
 $\langle r_m ^2\rangle_{I=1}$& $(1.12\, fm)^2$ & $(1.14\, fm)^2$ & $(0.80\, fm)^2$ \\[5pt] 
 $g_A$ & $1.27$ & $1.22$  & $1.26$ \\ [5pt]
 $\overline{N}_\pi$& $2.4$& $1.9$ & $\diagup$\\[10pt]
 \hline
\end{tabular} 
\end{table}

\begingroup
\begin{table}[b]
\caption{Various nucleon properties at Mean-Field level in the present model and in the $\sigma$-model~\cite{W.Broniowski1986} with vector mesons.}\label{resVM}
\centering
\begin{tabular}{|c | c  c  c|}
\hline
 & & & \\
 Quantity & Log. Model & $\sigma$-Model & Exp.  \\[7pt]
 \hline
  & & & \\
 $M\,(MeV)$ & $1329.5$ & $1331.7$ & $1085$\\[5pt]
 $\langle r_e ^2\rangle_{I=0}$& $(0.78\, fm)^2$ & $(0.76\, fm)^2$  & $(0.72\, fm)^2$ \\ [5pt]
 $\mu_{I=1}$ $(\mu_N)$& $4.49$ & $4.51$ & $4.70$ \\[5pt] 
 $\langle r_m ^2\rangle_{I=1}$& $(0.99\, fm)^2$ & $(1.01\, fm)^2$ & $(0.80\, fm)^2$ \\[5pt] 
 $g_A$ & $1.34$ & $1.35$  & $1.26$ \\ [5pt]
 $\overline{N}_\pi$& $2.62$& $2.66$ & $\diagup$\\[10pt]
 \hline
\end{tabular} 
\end{table} 
\endgroup

\begingroup
\begin{table}[t]
\caption{Contributions to the soliton total energy at Mean-Field 
level in the Logarithmic model and in the Linear $\sigma$-model~\cite{Birse1985}. All quantities are in MeV.}\label{contrib}
\centering
\begin{tabular}{|c c c|c|c|}
\hline
 & & & & \\
Quantity &  &  & Log. Model & Linear $\sigma$-Model  \\[7pt]
\hline
 & & & & \\
Quark eigenvalue &  &  & $83.1$ & $107.4$\\[5pt]
Quark kinetic energy &  &  & $1138.0$& $1056.9$ \\ [5pt] 
$E_\sigma$ (mass+kin.)&  &  & $334.5$& $320.3$ \\[5pt]
$E_\pi$ (mass+kin.)&   &  & $486.0$ & $373.1$\\ [5pt]
Potential energy $\sigma -\pi$& & & $105.7$& $120.7$ \\ [5pt]
$E_{q\sigma}$&   &   & $-101.4$ & $-62.3$\\ [5pt]
$E_{q\pi}$ &   &  & $-787.0$ & $-673.2$\\ [5pt]
Total energy  &   &  & $1175.6$ & $1136.2$\\ [10pt]  
 \hline
\end{tabular} 
\end{table}
\endgroup

\begingroup
\begin{table}[b]
\caption{Contributions to the soliton total energy at Mean-Field 
level in the Logarithmic model and in the Linear $\sigma$-model with vector mesons~\cite{W.Broniowski1986}. All quantities are in MeV.}\label{contribVM}
\centering
\begin{tabular}{|c c c|c|c|}
\hline
 & & & & \\
Quantity &  &  & Log. Model & Linear $\sigma$-Model  \\[7pt]
\hline
 & & & & \\
Quark eigenvalue &  &  & $114.5$ & $112.9$\\[5pt]
Quark kinetic energy &  &  & $1075.8$& $1080.6$ \\ [5pt] 
$E_\sigma$ (mass+kin.)&  &  & $213.8$& $212.2$ \\[5pt]
$E_\pi$ (mass+kin.)&   &  & $393.2$ & $397.3$\\ [5pt]
Potential energy $\sigma -\pi$& & & $81.2$& $80.4$ \\ [5pt]
$E_\omega$ (mass+kin.)&   &  & $-194.4$ & $-196.5$\\ [5pt]
$E_\rho$ (mass+kin.)&   &  & $162.6$ & $165.4$\\ [5pt]
$E_A$ (mass+kin.)&   &  & $329.5$ & $334.1$\\ [5pt]
$E_{q\sigma}$&   &   & $6.54$ & $4.74$\\ [5pt]
$E_{q\pi}$ &   &  & $-621.9$ & $-627.1$\\ [5pt]
$E_{q\omega}$ &   &  & $388.9$ & $393.0$\\ [5pt]
$E_{q\rho}$ &   &  & $-163.8$ & $-165.9$\\ [5pt]
$E_{qA}$ &   &  & $-341.8$ & $-346.4$\\ [5pt]
Total energy  &   &  & $1329.5$ & $1331.7$\\ [10pt]  
 \hline
\end{tabular} 
\end{table}
\endgroup

\begingroup
\begin{table}[t]
\caption{Projected nucleon properties in the present work 
without vector mesons and in the linear
$\sigma$-model and comparison with experimental values.}\label{resproj}
\centering
\begin{tabular}{|c | c   c  c|}
\hline
 & & & \\
 Quantity & Log. Model & $\sigma$-Model & Exp.  \\[7pt]
 \hline
  & & & \\
 $E_{1/2}\, (MeV)$ & $1075$ & $1002$ &  \\[5pt]
 $M_N \,(MeV)$ & $960$&  $ 894$& $938$ \\[7pt]
 $E_{3/2}\, (MeV)$ & $1140$ & $1075$ &  \\[5pt]
 $M_{\Delta}\, (MeV)$ & $1032$& $975$& $1232$\\[7pt]
  $\langle r_E ^2\rangle_p\, (fm^2)$ & $0.55$ & $0.61$ & $0.74$\\[5pt]
  $\langle r_E ^2\rangle_n\, (fm^2)$ & $-0.02$ & $-0.02$ & $-0.12$\\[5pt]
  $\langle r_M ^2\rangle_p\, (fm^2)$ & $0.7$ & $0.72$ & $0.74$\\[5pt]
 $\langle r_M ^2\rangle_n\, (fm^2)$ & $0.72$ & $0.75$ & $0.77$\\[5pt]
  $ \mu_p$ $(\mu_N)$ & $2.25$ & $2.27$ & $2.79$\\[5pt]
 $\mu_n$ $(\mu_N)$ & $-1.97$ & $-1.92$ & $-1.91$\\[5pt] 
 $g_a$ & $1.52$ & $1.10$ & $1.26$\\[5pt]  
                    & $1.6$ $(J=1/2)$& $1.2$ $(J=1/2)$ &  \\
  $\langle N_\pi \rangle_J$&                  &                  & $\diagup$\\
                    & $2.$ $(J=3/2)$& $1.6$ $(J=3/2)$& 	\\[10pt]
 \hline 
\end{tabular} 
\end{table} 
\endgroup

\begingroup
\begin{table}[b]
\caption{Projected nucleon properties  in the present work and in the linear
$\sigma$-model with vector mesons and comparison with experimental values.}\label{resprojVM}
\centering
\begin{tabular}{|c | c   c  c|}
\hline
 & & &  \\
 Quantity & Log. Model & $\sigma$-Model & Exp.  \\[7pt]
 \hline
  & & & \\
 $E_{1/2}\, (MeV)$ & $892$ & $882$ &  \\[5pt]
 $M_N \,(MeV)$ & $763$&  $ 750$& $938$ \\[7pt]
 $E_{3/2}\, (MeV)$ & $1030$ & $1029$ &  \\[5pt]
 $M_{\Delta}\, (MeV)$ & $918$& $917$& $1232$\\[7pt]
 $\langle r_E ^2\rangle_p\, (fm^2)$ & $0.59$ & $0.58$ & $0.74$\\[5pt]
 $\langle r_E ^2\rangle_n\, (fm^2)$ & $-0.03$ & $-0.02$ & $-0.12$\\[5pt]
  $\langle r_M ^2\rangle_p\, (fm^2)$ & $0.69$ & $0.69$ & $0.74$\\[5pt]
 $\langle r_M ^2\rangle_n\, (fm^2)$ & $0.70$ & $0.71$ & $0.77$\\[5pt]
  $ \mu_p$ $(\mu_N)$ & $2.72$ & $2.71$ & $2.79$\\[5pt]
 $\mu_n$ $(\mu_N)$ & $-2.49$ & $-2.5$ & $-1.91$\\[5pt] 
 $g_a$ & $1.6$ & $1.48$ & $1.26$\\[5pt]  
                    & $1.1$ $(J=1/2)$& $1.8$ $(J=1/2)$ &  \\
  $\langle N_\pi \rangle_J$&                  &                  & $\diagup$\\
                    & $1.3$ $(J=3/2)$& $2.2$ $(J=3/2)$& 	\\[10pt]
 \hline 
\end{tabular} 
\end{table} 
\endgroup

\begingroup
\begin{table}[t]
\caption{Projected nucleon properties  in the present work and in the linear
$\sigma$-model with vector mesons and comparison with experimental values for the parameter set: $g=3.6$, $g_\omega=13$, $g_\rho=4$ and $m_\sigma=1200$ MeV.}\label{resprojVMa}
\centering
\begin{tabular}{|c | c   c  c|}
\hline
 & & & \\
 Quantity & Log. Model & $\sigma$-Model & Exp.  \\[7pt]
 \hline
  & & & \\
   $E_{1/2}\, (MeV)$ & $1020$ & $1008$ &  \\[5pt]
 $M_N \,(MeV)$ & $926$&  $912$& $938$ \\[7pt]
 $E_{3/2}\, (MeV)$ & $1148$ & $1147$ &  \\[5pt]
 $M_{\Delta}\, (MeV)$ & $1066$& $1063$& $1232$\\[7pt]
 $\langle r_E ^2\rangle_p\, (fm^2)$ & $0.67$ & $0.66$ & $0.74$\\[5pt]
 $\langle r_E ^2\rangle_n\, (fm^2)$ & $-0.05$ & $-0.05$ & $-0.12$\\[5pt]
   $\langle r_M ^2\rangle_p\, (fm^2)$ & $0.77$ & $0.76$ & $0.74$\\[5pt]
 $\langle r_M ^2\rangle_n\, (fm^2)$ & $0.78$ & $0.77$ & $0.77$\\[5pt]
  $ \mu_p$ $(\mu_N)$ & $2.63$ & $2.64$ & $2.79$\\[5pt]
 $\mu_n$ $(\mu_N)$ & $-2.37$ & $-2.38$ & $-1.91$\\[5pt] 
 $g_a$ & $1.58$ & $1.46$ & $1.26$\\[10pt]  
 \hline 
\end{tabular} 
\end{table} 
\endgroup

In Tables~\ref{resproj} and~\ref{resprojVM} we present the results after projection in
both models, with and without vector mesons.
 Moreover,
in Table~\ref{resprojVMa} we present the results obtained
using a second parameter set, better fitted to the single nucleon properties.
It is important to remark that our results in the logarithmic model with only chiral fields
are consistent with the
ones obtained in~\cite{AbuShady:2010zza}. There, a different approach
based on the coherent pair approximation was used.
Their results are similar
to ours when the coherence length parameter $x$ is taken to be of the
order of one, as suggested in~\cite{PhysRevD.60.114022}. 

The results obtained both without and with the vector mesons 
in general overestimate the experimental values, particularly
for the magnetic observables,
 once the parameters are chosen so that
the projected mass of the nucleon is close to its physical value. One has anyway
to recall that for the mass an approximate correction for the spurious center
of mass motion has been taken into account (see eq.(~\ref{spurious})), while no
center of mass correction has been done for the other observables.
When this further corrections are taken into account the value of some
 observables typically reduces.

\section{Results: solitons at finite density}\label{hedgefindensity}
In this section we present the results obtained by studying a Wigner-Seitz
lattice of solitons.
We first discuss the energy of the system at finite density and we then present
the effect of the density on the single nucleon properties. 

\subsection{Energy of the lattice}
In Figs.~\ref{Etot} and~\ref{EtotVM}  we present the results of the total energy per unit cell in the present model and in the
linear $\sigma$-model with and without vector mesons.
  For each given value of $m_\sigma$, the
logarithmic model allows the system to reach higher densities. 
Notice that as
$m_\sigma$ increases, the system remains stable down to lower values of $R$ because the
chiral fields are more and more restricted to lay on the chiral
circle. Moreover the introduction of vector mesons stabilises even more
the solution, hence in comparison to the model with only $\sigma$ and $\pi$ we
can reach higher densities.

\begin{figure}[b]
\centering
\includegraphics*[width=\columnwidth]{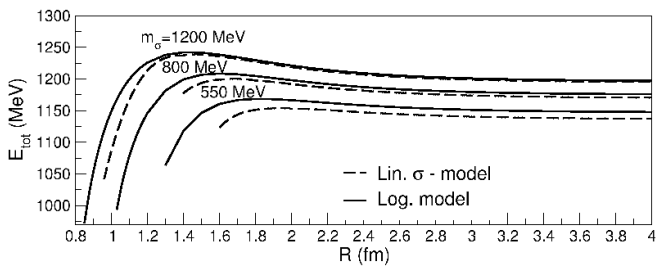} 
\caption{Total energy of the soliton as a function of cell radius $R$ for the linear 
$\sigma$-model \cite{U.Weber1998} and for the present model without vector mesons. 
Different values of $m_\sigma$ are considered, $g_\pi=5$, $g_\omega=12$ and $g_\rho=4$.}\label{Etot}
\end{figure}

\begin{figure}[t]
\centering
\includegraphics*[width=\columnwidth]{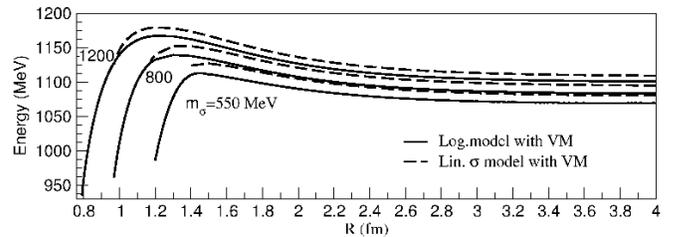} 
\caption{Total energy of the soliton as a function of cell radius $R$ for the linear 
$\sigma$-model \cite{U.Weber1998} and for the present model with vector mesons. 
Different values of $m_\sigma$ are considered.}\label{EtotVM}
\end{figure}

In Fig.~\ref{band} we plot the quark eigenvalue for the model without vector mesons
 as a function of the
cell radius $R$. The line labeled $\epsilon_{top}^{(a)}$ corresponds to the estimate
of the top of the band given by eq.~(\ref{bandglend}), while
$\epsilon_{top}^{(b)}$ follows from eq.~(\ref{bandmike}).
It is clear that in absence of vector mesons we never obtain saturation. Moreover
to change the value of $m_\sigma$ does not modify this result.
  In Fig.~\ref{bandVM} we show the analogous results in the case with vector mesons.
The band structure is similar to the one in Fig.~\ref{band}, but here the main difference 
is given by a significant increase of the top of the band
at high densities which allows us to obtain saturation.\\

\begin{figure}[b]
\centering
\includegraphics*[width=\columnwidth]{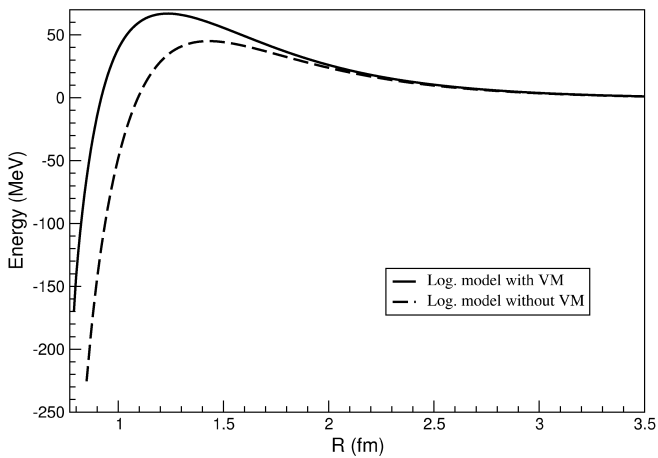} 
\caption{Total energy of the soliton at Mean-Field level in the logarithmic model without vector mesons (dashed line) and with vector mesons (solid line). The parameter values are: $m_\sigma=1200$ MeV, $g_\pi=5$, $g_\omega=12$ and $g_\rho =4$.}\label{FigEnTildateLOGVMNOVM}
\end{figure}

To better understand our result which indicates the possibility of getting
saturation, in Fig.~\ref{FigEnTildateLOGVMNOVM} we compare the total energy of the soliton at Mean-Field without 
the contribution associated with the band. In order to emphasize the effect of the density
on the energy we subtracted the mass of the nucleon in vacuum.
It is clear that the exchange of vector mesons plays a crucial role,
by contributing $\sim100$ MeV at $R=1$ fm, but it is not sufficient to get saturation.
To determine which ingredients of the model are actually providing the repulsion at high densities, in Fig.~\ref{entildate} we plot the interaction energies 
for each term contributing to the total energy at finite density. The plotted quantities are
 defined as the value of the chosen energy contribution
at $R$ minus the corresponding vacuum value:
\begin{equation}
\tilde{E}_i (R) = E_i(R)-E_i (\infty).
\end{equation}
From the figure it is clear that the band effect strongly influences the total energy of the soliton, by providing the largest contribution to repulsion at high densities. 
This is not surprising, because the band is associated with the sharing of quarks between nucleons.
It is well known~\cite{Shimizu:1989ye,Faessler:1996wt} that in calculations of the $N-N$ potential based on 
quark models the short-range repulsion is associated with the formation of a six-quark bag.
In our calculation the exchange of vector mesons is the dominant effect at densities up to
$\rho_0$, but at very high densities the band effect dominates.
The total amount of these contributions leads to the repulsive mechanism responsible for the steep rising of the total energy at high densities, as it can 
be seen in the upper panel of Fig.~\ref{bandVM}.\\

In Figs.~\ref{scomposizenVM} and~\ref{scomposizenVM1} we show in details all the contributions
to the total energy (without the band effect).
It is clear that the $\omega$ meson provides short-range repulsion, partially compensated
by the interaction of the quarks with the $A$ meson.

\begin{figure}[t]
\centering
\includegraphics*[width=\columnwidth]{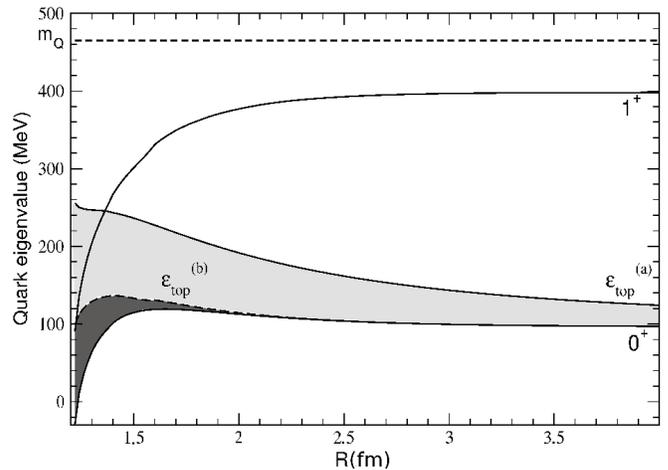} 
\caption{Quark eigenvalue as a function of the cell radius $R$,
in the model without vector mesons. 
The shaded areas represent the band as estimated in eq.(\ref{bandglend}) and in eq.~(\ref{bandmike}). 
The first excited state $1^+$ is also shown. The quark mass in vacuum, here $465$ MeV, 
is indicated by the dashed line.}\label{band}
\end{figure}

\begin{figure}[b]
\centering
\includegraphics*[width=\columnwidth]{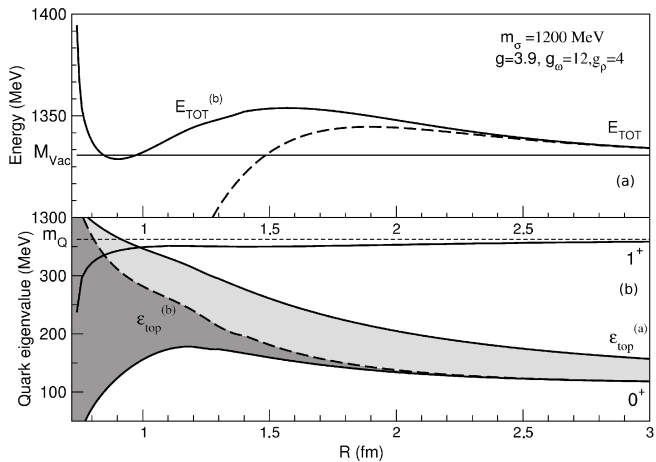} 
\caption{Panel (a): total energy of the soliton with band effects (solid line) and without band effects (dashed line) as a function of the cell radius $R$ for the model with vector mesons. Panel (b): the quark eigenvalue as a function of the cell radius $R$ for the model with vector mesons. 
The shaded areas represent the band as estimated in~ eq.(\ref{bandglend}) and in eq.~(\ref{bandmike}). 
The quark mass, here $362.7$ MeV, 
is indicated by the dashed line. }\label{bandVM}
\end{figure}

\begin{figure}[t]
\centering
\includegraphics*[width=\columnwidth]{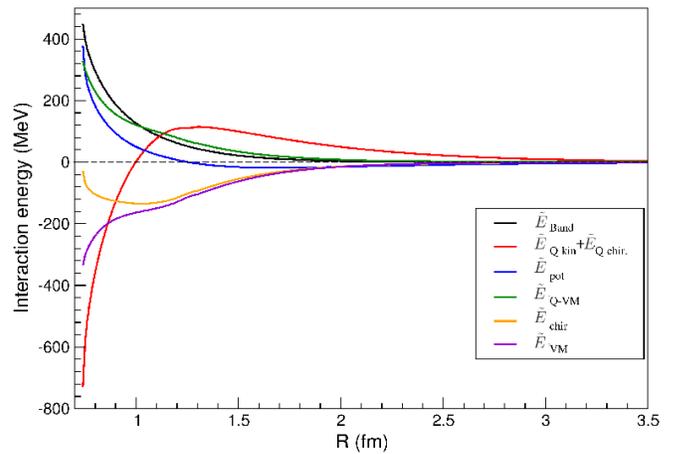} 
\caption{Color online. Interaction energies $\tilde{E}_i$, as defined in text,  as a function of the cell radius $R$ in the model with vector mesons. Parameters as in Fig.~\ref{bandVM}.}\label{entildate}
\end{figure}

\begin{figure}[b]
\centering
\includegraphics*[width=\columnwidth]{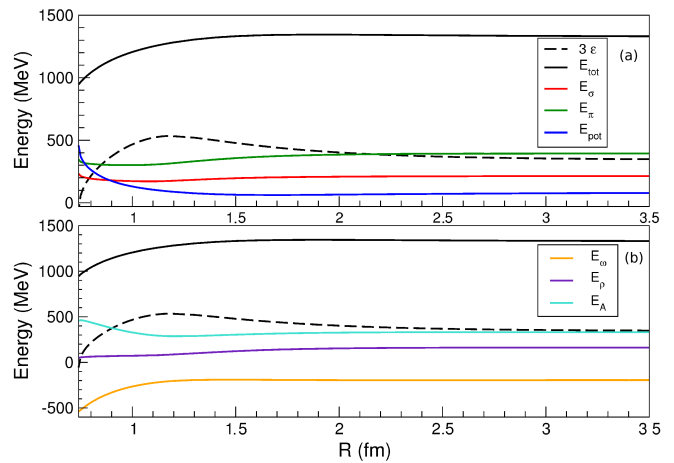} 
\caption{Color online. Panel (a): contributions of the chiral fields to the total energy of the soliton as a function of the cell radius $R$ in the model with vector mesons. Panel (b): contributions of the vector meson fields to the total energy of the soliton as a function of the cell radius $R$. Parameters as in Fig.~\ref{bandVM}.}\label{scomposizenVM}
\end{figure}

\begin{figure}[t]
\centering
\includegraphics*[width=\columnwidth]{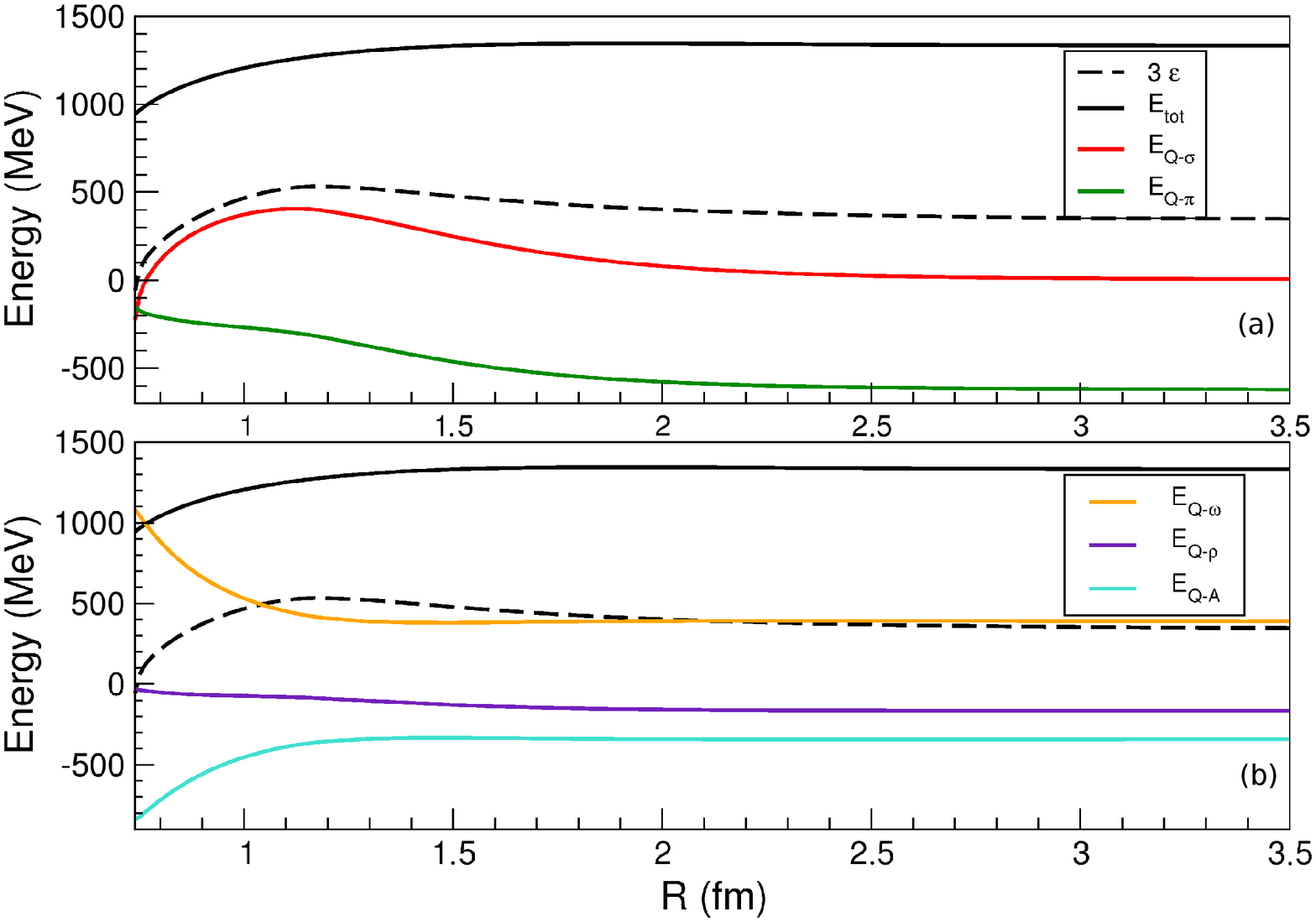} 
\caption{Color online. Panel (a): interaction energies of the chiral fields with quarks as a function of the cell radius $R$ in the model with vector mesons. Panel (b): interaction energies with the vector mesons as a function of the cell radius $R$. Parameters as in Fig.~\ref{bandVM}.}\label{scomposizenVM1}
\end{figure}
It has been discussed in the literature how to interpret the results obtained
using the Wigner-Seitz lattice and in particular which should be the
indications of quark deconfinement. In~\cite{D.Hahn1987} it has been
suggested that deconfinement takes place when the upper band, which
corresponds to Grand Spin $G=1$, merges with the lower
band.
In the case with vector mesons this occurs roughly at densities
slightly larger than the saturation density.
We should keep in mind that the estimate of the width of the band is affected by large uncertainties
and it is well possible that in a more refined calculation saturation density and
deconfinement density turn out to be well separated.

Which conclusions can be drawn from our analysis?
First and foremost we have a scenario in which it is possible, for the first time,
to obtain saturation by making use of the rather common idea of getting attraction
from the chiral fields and repulsion from the vector fields. In this game the
logarithmic potential plays a crucial role, by allowing the solitons to remain
stable at densities large enough that the vector mesons can play a role.
It is important at this point to clarify whether we can get this result only
in a special and tiny parameters' range or if the mechanism leading to saturation
is rather stable respect to the choice of the parameters' values. 
This analysis is shown in Fig.~\ref{coester} where we plot the value of the total energy,
including the band effects, at the "saturation density" point minus the energy in
vacuum for different values of the parameters, at fixed $m_\sigma=1200$ MeV.
Here "saturation density" means the density at which a local minimum in the total energy
appears, even if that minimum is not the global one. The minimum is global when the 
energy plotted in Fig.~\ref{coester} is negative and in that situation we are getting
real saturation. Instead, when the plotted energy is positive the local minimum corresponds
to a sort of metastable state. Since our calculation is affected by large uncertainties
we think it is interesting to show also the parameters leading to this "false" minimum,
because in a more sophisticated calculation (based for instance on a better estimate of the
band) those energies could easily become negative. For parameters' values significantly
outside the indicated range no local minimum exists. For instance for values of $g_\omega > 12$
the local minimum of the energy disappears because the energy keeps raising as the density increases, while for small
values of $g_\omega$ the repulsion cannot contrast the attraction and the energy gets smaller and smaller
at large densities. 
It is important to notice that the range of parameters' values providing "saturation" is 
at least in part overlapping
with the range of parameters' values for which a reasonable description of the single soliton in vacuum can 
be obtained. Finally, we have to recall that we are not really studying nuclear matter, but rather
$G=0$ matter, which is composed of degenerate nucleons and deltas. Obviously, no experimental data
exist for that type of theoretical matter, but we can expect it to be saturating, probably with a larger
saturation energy than that of nuclear matter. The corresponding saturation density is also obviously
unknown.

\begin{figure}[t]
\centering
\includegraphics*[width=\columnwidth]{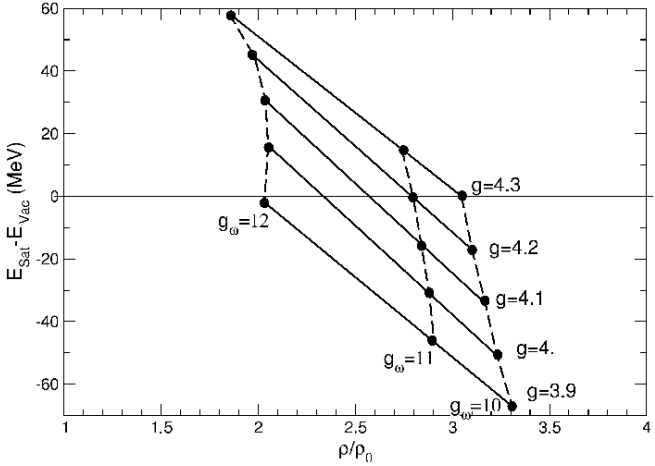}
\caption{Minimum of the total energy as a function of the density. Different sets of couplings are shown, for which the model
admit saturation. See the text for more details.} \label{coester}
\end{figure}

\subsection{Nucleon properties at finite density}\label{properties}
The question about the
modification of the nucleon properties at finite density have been
investigated in many analysis, both experimental~\cite{Aubert:1983xm}
and theoretical~\cite{PhysRevLett.53.892,Ericson:1986ud,Mulders:1990xw,Lu:1998tn,PhysRevC.70.065205}.
In Figs.~\ref{obs} and~\ref{obsVM} we show how the values of
a few observables evolve as a function of the spherical cell density
$\rho_C=(4 \Pi R^3/3)^{-1}$. 
The Mean-Field observables evaluated as a function of the density cell are the isoscalar electric and the isovector magnetic radius and the number of pions.
The formulae for these quantities for the model without and with vector mesons read~\cite{Cohen:1986va,Ruiz1995}:
\begin{align}\label{obsmeanf}
& \langle r^2_E\rangle_{I=0}=\int r^4 (u^2+v^2)dr\nonumber\\
& \langle r^2 _M\rangle_{I=1}=\frac{1}{\mu_{I=1}}\int r^5 \frac{2\pi}{9}(G_{M} ^Q+G_{M} ^\pi +G_{M} ^\rho +G_{M} ^A) \,,
\end{align}   
and $\mu_{I=1}$ and the radial functions $G_m ^{field}$ are given in Appendix III. 
The problem with our evaluation is that in the Wigner-Seitz approach
we are forcing a unit of baryon number in every cell of the lattice.
Therefore, at densities large enough that the fields start occupying most of the cell and their
value is no more strongly varying inside the cell, the various radii all simply scale
with the size of the cell. The results we are obtaining are therefore indicative
only at densities low enough that the fields are still relatively well
contained inside each cell.
As shown in Figs.~\ref{campi} and ~\ref{campiVM} all the fields are well confined up to values of density close to $\rho_0$.
This implies that the behaviour
of the observables, evaluated from these fields,  have no physical relevance at densities
of the order or above nuclear matter saturation density $\rho_0$.

The problem we are facing is deeply associated with the Wigner-Seitz approximation
in which the effect of the finite density to the various observables is due only to Hartree contributions. In the real case two neighboring nucleons interact also via the Fock term.
Notice that at densities large enough that the field fluctuations are suppressed, the Hartree
contributions associated e.g with the pion field vanishes.
On the contrary the Fock term becomes relevant at those densities and it provides 
contributions to the electromagnetic
observables which in nuclear physics are sometimes called pion-in-flight.
These terms, of course, cannot be evaluated in the Wigner-Seitz approach. 

 Another problem with the Wigner-Seitz approach is that
it imposes spherical boundary conditions on the fields. This is particularly
dangerous in the case of the chiral fields, since at Mean-Field level directions
in ordinary space are connected with directions in isospin space, a situation
which is certainly quite far from reality. Due to these problems a work is in
progress~\cite{Sarti:2012un} in which a real lattice will be studied, with
boundary conditions which can change depending on the direction.
We think it is nevertheless worthy to present our results in Figs.~\ref{obs}-\ref{obsVM} 
so to compare them with future more precise estimates.

For the model containing $\sigma$ and $\pi$ only, our results can be meaningful at low densities where
the dynamics is dominated by chiral fields.
The introduction of vector mesons affects in quite a interesting way the isoscalar radius: as shown in Fig.~\ref{densityvm}, the repulsion provided by the $\omega$ field, for densities smaller than $\rho_0$, prevents the swelling of the nucleons. 
The qualitative effect of the inclusion of the vector mesons is to stabilize the shape of the solitons respect to compression.
This can be seen also in the case of the magnetic radius where the reduction of this quantity as a function of the density is less marked than in the case without vector mesons.
It is also possible to evaluate the so-called "super-ratio" defined as $(G_E/G_M)^\rho/(G_E/G_M)^{vac}$,
where $G_{E,M}$ are the electric and magnetic form factors.
In our calculation we obtain a reduction of this quantity as a function of the density, similarly to what obtained in other works, although the effect here is much larger. Let us stress again that
although the model with vector mesons allows to reach much larger densities, we do not 
attribute too much significance to the behaviour of the observables at $\rho\gtrsim \rho_0$ for the reasons explained above.
In the lowest panel of Figs.~\ref{obs} and~\ref{obsVM}
we also show the behaviour of the number of
pions $\overline{N}_\pi$ per unit cell at finite density. The possibility
of an enhancement of the pion cloud, when the nucleon is not isolated, has been
discussed in the literature~\cite{Ericson:1986ud}.  In the present approach we obtain instead a
decrease of $\overline{N}_\pi$.
This result stems from the behaviour of the pionic field on the Wigner-Seitz lattice as shown in Fig.~\ref{campi}. The strong reduction of the pionic field is due to the boundary conditions
requested by the spherical symmetry.
It will be interesting to see if this result survives when a real lattice is used in the calculation.
\begin{figure}[t]
\centering
\includegraphics*[width=\columnwidth]{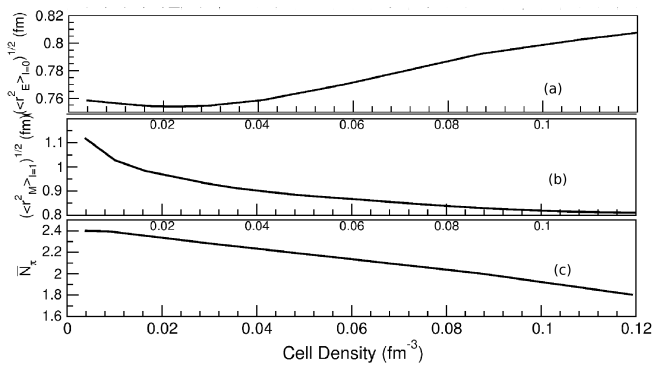}
\caption{Isoscalar electric radius (panel (a)), isovector magnetic radius (panel (b))
and average number of pions (panel (c)) as a function of cell density $\rho_C$ for the model without
vector mesons.  
}\label{obs}
\end{figure}

\begin{figure}[t]
\centering
\includegraphics*[width=\columnwidth]{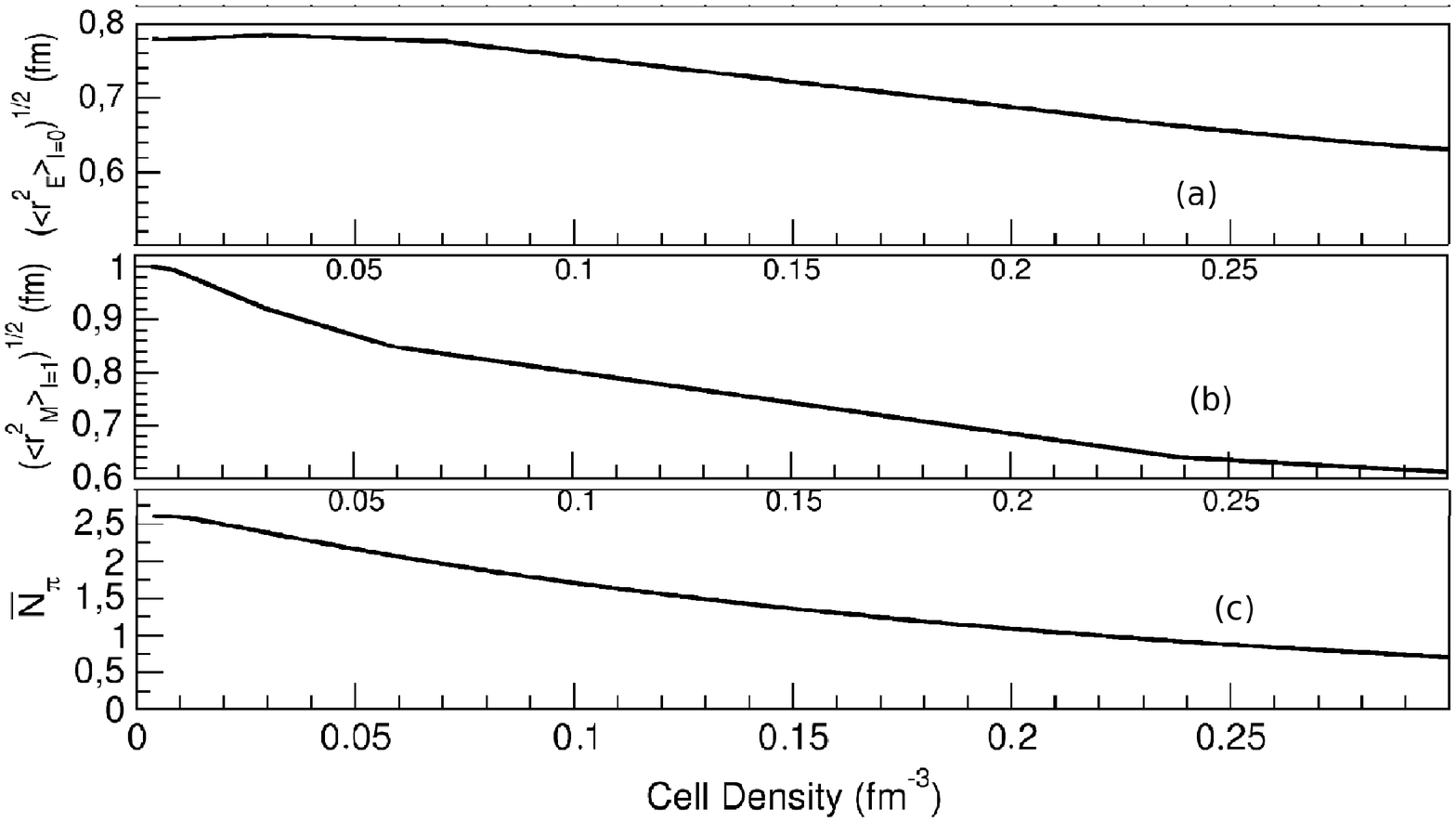}
\caption{Isoscalar electric radius (panel (a)), isovector magnetic radius (panel (b))
and average number of pions (panel (c)) as a function of cell density $\rho_C$ for the model with vector mesons.}\label{obsVM}
\end{figure}

\section{Conclusions}\label{conclusions}
We used a Lagrangian with quarks degrees of freedom based on chiral and scale 
invariance to study how the soliton behaves in vacuum and at finite density.
We presented results for the simple model with just chiral fields and also for the
model including vector mesons.
To describe the single nucleon properties in vacuum we have
used a projection technique.
The values of the static observables in the
logarithmic model at zero density are comparable to the ones
obtained using the linear $\sigma$-model.
  
For the description of
the soliton at finite density we have employed the Wigner-Seitz
approximation.
We have shown that the new potential, which includes 
a logarithmic term originating from the breaking of scale invariance,
allows the system, for each given $m_{\sigma}$, to reach
densities larger than the ones obtained with the $\sigma$-model.
Moreover, as expected, the addition of the vector mesons plays a
double role in the study at finite density: on one hand it stabilizes the solution
and allows to reach even higher densities, on the other hand it partially provides the 
repulsion necessary to obtain saturation. 
The remaining repulsive contribution originates from the band effect.

The possibility to obtain saturation seems to be a firm result of the model, at the level of the Wigner-Seitz approximation.
In fact, by exploring the space of parameters, we have shown that the interplay between 
attraction, provided by the chiral fields, and repulsion, given by the omega field and the band effect,
allows to get saturation for a rather extended range of parameters' values.

At sub-nuclear densities the dynamics should be dominated by
the chiral fields and the modifications
of the nucleon observables obtained in our work can
therefore be physically relevant in the low-density range.
In particular the
isoscalar radius presents a slight swelling, of the order of $5\%$
in the model without vector mesons and at even smaller dependence on the density in the complete
model. 
This trend is in agreement with previous calculations~\cite{PhysRevLett.53.892,Lu:1998tn,Smith:2004dn}.

The present work will be extended in several directions. First
 a more precise and accurate
calculation of the band in the soliton crystal will be 
done following Ref.~\cite{U.Weber1998}.  
Work is in progress in order to study this same model by using the 
technique developed in Ref.~\cite{Park:2008xm}, which provides
a more precise description of a multi-soliton system.
Finally, the
model can also be studied at finite temperature, including the
dynamics of the dilaton field. 
We can expect that the effect of the finite temperature
on the soliton lattice will be to reduce the stability, by lowering
the value of the dilaton field and therefore making it more easy
for the chiral fields to fluctuate. It will be interesting
to compare the obtained phase diagram 
with the one proposed by McLerran and Pisarski~\cite{L.McLerran2007}. 

\begin{acknowledgements}
It is a pleasure to thank B.Y.Park and V. Vento for many stimulating discussions, 
M.Birse and J.McGovern for useful comments and tips on calculations.\\
\end{acknowledgements}

\appendix* 
\section*{APPENDIX I}\label{appendix1}
In this Appendix we provide the explicit expression for the energy density at Mean-Field given in eq.~(\ref{enmfa}).\\
The quark-mesons interaction and the quark kinetic energies are:
\begin{eqnarray}
 E_{int}&=&\dfrac{3}{4\pi} \left\lbrace g_\pi   \sigma_h (u^2-v^2)+2g_\pi h uv +\dfrac{g_\omega}{3}\omega (u^2+v^2) \right.\nonumber\\
 & & \left.   -2g_\rho \rho uv +g_\rho \left[\dfrac{3}{2}A_S \left( u^2-\dfrac{1}{3}v^2 \right) +\dfrac{2}{3}v^2A_T\right] \right \rbrace \nonumber \\
\end{eqnarray}
\begin{align}
& E_{kin,Q}= \dfrac{3}{4\pi} \left(u \dfrac{dv}{dr}-v \dfrac{du}{dr}+\dfrac{2}{r}uv\right)
\end{align}
and the energy density of the mesons fields and of the potential read:
\begin{align}
E_{\sigma}=& \dfrac{\beta}{2}\left[ -\dfrac{d\sigma_h}{dr}-g_\rho h \left(A_S+\dfrac{2}{3}A_T\right)\right]^2\\
E_\pi =&\dfrac{\beta}{2}\left[\dfrac{dh}{dr}+g_\rho \sigma_h \left(A_S +\dfrac{2}{3}A_T\right)\right]^2\nonumber\\
 & +\beta  \left[ -\dfrac{h}{r}+g_\rho h +g_\rho \sigma_h \left(A_S -\dfrac{1}{3}\right)\right]^2\\
E_\omega =& -\dfrac{1}{2}\left(\dfrac{d\omega}{dr}\right)^2 -\dfrac{1}{2}m_\omega ^2 \omega ^2\\
E_\rho =& \left[ \dfrac{d\rho}{dr}+\dfrac{\rho}{r}-g_\rho \left(A_S +\dfrac{2}{3}A_T\right)\left(A_S -\dfrac{1}{3}A_T\right)\right]^2\nonumber\\
&  +\dfrac{1}{2}\left[ \dfrac{2}{r}\rho -g_\rho \rho ^2 -g_\rho \left(A_S -\dfrac{1}{3}A_T\right)^2\right]^2 +m_\rho ^2 \rho^2 \nonumber\\
&\\
E_{A}=& \left[ \left( \dfrac{dA_S}{dr}-\dfrac{1}{3}\dfrac{dA_T}{dr}\right)-\dfrac{A_T}{r}+g_\rho \rho \left(A_S +\dfrac{2}{3}A_T\right)\right]^2\nonumber\\
&  +\dfrac{1}{2}m_\rho ^2\left(3A_S^2 +\dfrac{2}{3}A_T^2\right)\\
E_{pot}=&V(\phi_0 ,\sigma_h , h)
\end{align}

\section*{APPENDIX II}\label{appendix}
In this Appendix we provide a detailed calculation of the expectation
value of the logarithmic potential given in eq.~(\ref{potfro}) between
the projected states (see eq.~(\ref{projen})).  As already mentioned
and shown in~\cite{Birse:1986qc}, terms which do not involve the pion
field (such as the quark-pion interaction energy and the $\sigma$ and
quark kinetic energies) are not affected by projection.  The main
issue is the evaluation of the matrix elements of the chiral fields
between rotated and unrotated hedgehog states.\\ These matrix elements
for the sigma field $\sigma(\boldsymbol{r})$ are:
\begin{align}\label{matelscal}
& \langle B\vert \widehat{R}(\Omega)^{-1} \sigma(\boldsymbol{r})^n \vert B \rangle = 
\overline{\sigma}(\boldsymbol{r})^n \langle B\vert  \widehat{R}(\Omega)^{-1} \vert B \rangle\, ,
\end{align}
where:
\begin{align}
& \overline{\sigma}(\boldsymbol{r}) =\dfrac{1}{2}\left(\sigma(\boldsymbol{r})+ 
\widehat{R}(\Omega)^{-1}\sigma(\boldsymbol{r})\right)\equiv \sigma_h (r)\, .
\end{align}
In an analogous way, for the pion field the matrix elements become:
\begin{align}
& \langle B\vert  \widehat{R}(\Omega)^{-1} \boldsymbol{\pi}(\boldsymbol{r}) \vert B \rangle = 
\overline{\boldsymbol{\pi}}(\boldsymbol{r})  h(r)\langle B\vert  \widehat{R}(\Omega)^{-1} \vert B \rangle\, ,\\
& \overline{\boldsymbol{\pi}}(\boldsymbol{r})= \dfrac{1}{2}\left(\widehat{\boldsymbol{r}}+
\widehat{R}(\Omega)^{-1}\widehat{\boldsymbol{r}}\right)\, .
\end{align}
Since the potential is a function of the pion only through quadratic terms, by using the previous relation we get:
\begin{align}\label{gfunc}
&\langle B\vert  \widehat{R}(\Omega)^{-1} \boldsymbol{\pi}^2(\boldsymbol{r}) \vert B \rangle =
g(\Omega,\theta,\phi) \langle B\vert  \widehat{R}(\Omega)^{-1} \vert B \rangle\nonumber\, ,\\
& g(\Omega,\theta,\phi)= 
\dfrac{1}{2} h(r)^2 \left(1+\widehat{\boldsymbol{r}}\cdot \widehat{R}(\Omega)^{-1}\widehat{\boldsymbol{r}}\right)\, ,
\end{align}
where the function $g$ depends on Euler angles $\Omega$ and on the polar and azimuthal angles.\\
For a generic function $F$ of the quadratic pionic terms, the following relation holds:
\begin{equation}
\langle B\vert  \widehat{R}(\Omega)^{-1} F[\boldsymbol{\pi}^2(\boldsymbol{r})] \vert B \rangle = 
F[g(\Omega,\theta,\phi)]\langle B\vert  \widehat{R}(\Omega)^{-1} \vert B \rangle.
\end{equation}
Therefore the projection of the potential term can be obtained by
leaving the pure $\sigma$ terms unchanged and by replacing the quadratic
terms of the pion with the function $g$ given in eq.~(\ref{gfunc}):
\begin{align} \label{potproj}
 V(\sigma_h ,& h, g(\Omega,\theta,\phi))  =\nonumber\\
& \lambda_1 ^2 \left(\sigma_h ^2 +h^2 g(\Omega,\theta,\phi)\right)-\lambda_2 ^2\,\ln\left(\sigma_h ^2 + 
h^2 g(\Omega,\theta,\phi)\right)\nonumber\\
& -\sigma_0 m_\pi ^2 \sigma_h\, .
\end{align}
The expectation value of the potential between the projected states, eq.~(\ref{potproj}), becomes:
\begin{align}
E_{J,pot}& = \langle  JJ-J\vert   :\int d^3r V(\sigma_h,h,  g(\Omega,\theta,\phi)):\vert JJ-J\rangle \nonumber\\
& =\dfrac{1}{N_J}\int _0 ^\infty r^2 dr \int _0 ^\pi \sin\theta d\theta \int_0 ^{2\pi} d\phi \nonumber\\
& \times \int d\Omega^3 D_{J,J}^J (\Omega) V(\sigma_h,h, g(\Omega,\theta,\phi)) 
\langle B\vert \widehat{R}(\Omega)^{-1} \vert B \rangle
\end{align}
where the Wigner function is equal to:
\begin{equation}
D_{J,J}^J (\Omega) =e^{-iJ(\alpha+\gamma)} \left(\cos\dfrac{\beta}{2}\right)^{2J}\, .
\end{equation}
Finally, the overlap between rotated and unrotated hedgehog states reads:
\begin{align}
\langle B\vert & \widehat{R}(\Omega)^{-1} \vert B \rangle =\nonumber\\
& \left(\cos \dfrac{\beta}{2} \cos \dfrac{\alpha+\gamma}{2}\right)^3\nonumber\\
& \times \exp \left( \overline{N}_\sigma +\dfrac{\overline{N}_\pi}{3}\left(4\cos^2 \dfrac{\beta}{2} 
\cos^2 \dfrac{\alpha+\gamma}{2}-1\right)\right).
\end{align}
Here $\overline{N}_\sigma$, $\overline{N}_\pi$ are the average numbers
of $\sigma$ and $\pi$ mesons in the hedgehog state and $N_J$ is a
normalization integral; explicit expressions of these quantities can
be found in~\cite{Birse:1986qc,Ruiz1995}.

\section*{APPENDIX III}\label{appendixIII}
The Mean-Field approximation for the isovector magnetic moment reads:
\begin{align}
&\mu_{I=1}=\int r^3 dr\frac{2\pi}{9}(G_{m} ^Q(r)+G_{m} ^\pi (r)
 +G_{m} ^\rho (r)+G_{m} ^A(r))\, ,
\end{align}
where the radial functions $G_{i} ^{field}$ are given by:
\begin{align}\label{contribobsmeanfield}
 G_M ^Q(r) = & \frac{3}{\pi} u v \,\,\, ,\nonumber\\
 G_M ^\pi(r) =&  4\beta \left[ \frac{1}{r} h^2 -g_\rho h \sigma_h \left(A_S -\frac{1}{3}A_T\right)-g_\rho h^2 \rho \right] \,\,\, ,\nonumber\\
 G_M ^\rho(r) =&4 \left\lbrace \frac{2}{r}\rho ^2 -g_\rho \rho \left[ \rho^2 +\left(A_S-\frac{1}{3}A_T\right)^2\right]\right\rbrace  \,\,\, ,\nonumber\\
 G_M ^A (r)= &(-1)\left[4 \left( A_S '-\frac{1}{3}A_T '-\frac{A_T}{r}\right) \left( A_S +\frac{2}{3}A_T\right) \right.\nonumber\\
& \left. +4 g_\rho \rho \left(A_S+\frac{2}{3}A_T\right)^2\right] \,\,\, .
\end{align}
For the model without vector mesons, the coupling constant $g_\rho$ and the meson fields vanish.
\bibliography{biblio}
\bibliographystyle{apsrev4-1}

\end{document}